\documentclass{IEEEtran4PSCC}

\usepackage{cite}

\ifCLASSINFOpdf
   \usepackage[pdftex]{graphicx}
\else
   \usepackage[dvips]{graphicx}
\fi
\usepackage[cmex10]{amsmath}
\usepackage{array}

\ifCLASSOPTIONcompsoc
  \usepackage[caption=false,font=normalsize,labelfont=sf,textfont=sf]{subfig}
\else
  \usepackage[caption=false,font=footnotesize]{subfig}
\fi

\usepackage{wasysym}
\usepackage{xcolor}
\usepackage{bbm}
\usepackage{soul,color} 

\makeatletter
\let\old@ps@headings\ps@headings
\let\old@ps@IEEEtitlepagestyle\ps@IEEEtitlepagestyle
\def\psccfooter#1{%
    \def\ps@headings{%
        \old@ps@headings%
        \def\@oddfoot{\strut\hfill#1\hfill\strut}%
        \def\@evenfoot{\strut\hfill#1\hfill\strut}%
    }%
    \def\ps@IEEEtitlepagestyle{%
        \old@ps@IEEEtitlepagestyle%
        \def\@oddfoot{\strut\hfill#1\hfill\strut}%
        \def\@evenfoot{\strut\hfill#1\hfill\strut}%
    }%
    \ps@headings%
}
\makeatother

\begin{document}
%
\title{Multi-Agent Learning in Double-side Auctions for Peer-to-peer Energy Trading}

\author{
\IEEEauthorblockN{Zibo Zhao\\ Andrew L. Liu}
\IEEEauthorblockA{School of Industrial Engineering \\
Purdue University\\
West Lafayette, IN, USA}
\{zhao438,  andrewliu\}@purdue.edu
}


\maketitle

\begin{abstract}
Distributed energy resources (DERs), such as rooftop solar panels, are growing rapidly and are reshaping power systems. To promote DERs, feed-in-tariff (FIT) is usually adopted by utilities to pay DER owners certain fixed rates for supplying energy to the grid. An alternative to FIT is a market based approach; i.e., 
consumers and DER owners trade energy in an auction-based peer-to-peer (P2P) market, and the rates are determined by a market clearing process. However, the complexities in sucha market and agents' bounded rationality may invalidate many well-established theories on auction design and hinder market development. To address this issue, 
we propose an automated bidding framework in a repeated auction based on multi-armed bandit learning, which aims to minimize each bidder's cumulative regret. Numerical results indicate convergence of such a  multi-agent learning game to a steady-state. For comparison purpose, we apply the framework to three different auction designs to realize a P2P market.
\end{abstract}



\section{Introduction}
\label{sec:intro}
Distributed energy resources (DERs) are a vital part of a smart grid, as such resources can improve system resilience with their proximity to load and promote sustainability, with the majority of DERs being solar and wind resources \cite{jiayi2008review, akorede2010distributed}. To incentivize investments in DERs, there are two general approaches: non-market-based versus market-based. 
The most common and widely used policy in a non-market-based approach is feed-in-tariff (FIT) \cite{lesser2008design} (including net-metering). While effective in promoting DERs, it may create equity issues as consumers without DERs would face increased electricity rates to pay for the FIT. In a market-based approach, a marketplace exists for consumers and DER owners, also referred to as prosumers, to trade energy among themselves. The rates that market participants pay/receive would fluctuate over time, reflecting the dynamic supply and demand conditions. 

In a bilateral marketplace, a leading mechanism to match supply and demand is through a double-side auction.  
While auction designs have been well studied in the field of economics and game theory \cite{friedman2018double, huang2002design, niu2013maximizing, nicolaisen2001market}, several special features of a peer-to-peer (P2P) energy market require special attention. To name a few,  
a P2P energy market inherently involves repeated auctions and exogenous uncertainties (e.g., wind/solar availability), making the analysis of market participants' bidding/asking strategies much more difficult. In addition, market participants are likely to have bounded rationality in the sense that they may not even know their own valuation of energy production and consumption (such as how much `utility' it would provide a consumer if he/she does dish washing at 8 PM versus at 11 PM). 
Furthermore, their (implicit) valuations are likely dependent. For example, in a hot summer day, all buyers would value high of energy consumption for air conditioning. 
This feature alone would nullify the assumptions of most of the works in auction theory. 

To address the theoretical difficulties, and to provide an algorithmic-framework that can be automated to aid consumers and prosumers to participate in a repeated double-auction, we propose a multi-agent, multi-armed bandit learning approach. More specifically, consider a double-auction in which buyers/sellers submit bids/asks in each time period $h$ (e.g., hourly) in each day $d$. Within each $h$, we assume that market participants only choose a price to ask/bid, not quantities of energy. We further discretize per-unit price bids (i.e., $\cent$/KWh) into $K$ possible choices. 
When each agent decides which price to bid/ask, it is similar to choosing one slot machine to pull the arm, out of $K$ such machines. In this case, the agents are uncertain if they will win (bids cleared) or lose (bids not cleared), and in the case of winning, how much the payoff would be. This is similar to the classic multi-armed bandit (MAB) learning problem that has been well studied in the  literature, such as \cite{lai1985asymptotically, auer2002finite, powell2012optimal, mahajan2008multi}. A key difference here, however, is that an  agent's probability of winning and the payoff distributions (of each arm) depend on how other market participants bid/ask, which give rise to a multi-agent MAB game. In such a game, each agent solves their own decision-making problem by choosing a learning algorithm. Many such algorithms have been proposed in the literature, such as the $\epsilon$-greedy and variants of the Upper Confidence Bound (UCB) algorithm\cite{auer2002finite}. A salient feature of the mutli-agent MAB game is that it does not require all the agents to use the same algorithm, nor does it require the agents to be homogeneous (in terms of their preferences and feasible action spaces).

We apply the mutli-agent MAB game framework and run multiple simulations to study three different auction designs, all of which could be used to organize a P2P energy trading market. The three auction designs are a replicate of the wholesale market's uniform-price auction, a variant of Vickrey double-side auction \cite{huang2002design}, and maximum volume matching auction (which is pay-as-bid/receive-as-ask) \cite{niu2013maximizing}. Numerical results indicate the convergence of the market outcomes of a MAB game to a steady-state. Based on the simulations, from market participants' perspective, the uniform-price auction outperforms the other two as it can offer higher cleared quantities, total social welfare and total (normalized) rewards.

The rest of the paper is structured as follows. In Section \ref{sec:mab-game}, we describe in details how market participants bid/ask through bandit learning in a double-side auction. In Section \ref{sec:double_auction}, three specific double-side auction designs to realize a P2P energy trading market are presented. Numerical simulations are shown in Section \ref{sec:num_res} by comparing learning results in the three different auction mechanisms. Section \ref{sec:conclusion} concludes the paper and identifies potential future research directions.

\section{Learning under MAB-game Framework}
\label{sec:mab-game}
Without a P2P energy market, DER owners can only sell their generated energy to a  utility company or an aggregator at some pre-defined fixed FIT; similarly, they buy energy from a utility at some fixed rate. With a bilateral P2P marketplace, consumers and prosumers can trade with each other at rates accepted by both sides. Intuitively, such a marketplace is desirable to all participants if it can provide sellers with higher-than-FIT rates and buyers lower rates than the utility's fixed rate. A double-side auction is arguably the most common approach to organize a P2P market. While buyers/sellers in such an auction can bid both price and quantities, as a starting point, for the energy auction, we assume that participants only bid prices.\footnote{While market participants do not bid quantities in our study, the quantities are not fixed however. They are stochastic as they are assumed to be from wind/solar generation.}

To help buyers/sellers decide what prices to bid/ask in a repeated auction setting, we propose a regret-minimization-based learning approach for a multi-agent system in which bidding/asking prices of agents are automatically chosen by bandit learning algorithms. 
For the ease of argument, we present the model setup corresponding to a single trading-period $h$ (e.g. 1 hour), and the same bidding process is repeated in every $h$. Thus we do not put an $h$ index for all the variables/parameters in the discussion. Consider a set of agents $\mathcal{A} = \mathcal{A}_b \cup \mathcal{A}_s$, where $\mathcal{A}_b$ and $\mathcal{A}_s$ are the sets of buyers and sellers, respectively. Further, we let $P_{FIT}$ denote the FIT rate, and $\overline{P}$ be the electricity rate charged by a utility, both in $\cent$/KWh. 
It is reasonable to assume that $P_{FIT} < \overline{P}$ (cause otherwise, utilities will be losing money, which we assume to be not reasonable, especially to investor-owned utilities).

\subsection{Discrete Price Arms}
\label{subsec:price_arms}
The majority of DERs are solar and wind resources,\footnote{As a starting point, we do not consider energy storage, and discuss the challenges of including storage in Section \ref{sec:conclusion}.} and thus we consider their generation marginal costs as zero, despite their fixed investment and installation fees (which are sunk costs). Therefore, getting paid at any rate higher than FIT for the energy they sell would be attractive to DER owners. Similarly, energy buyers prefer any rate lower than the utility rate. Therefore, any rate in the range $[P_{FIT}, \overline{P}]$ would be preferred to both buyers and sellers. Let $\mathcal{P}_i$ denote the feasible bidding/asking price space for each agent $i$; then we have $[P_{FIT}, \overline{P}] \subset \mathcal{P}_i$. Note that it is not the other way around since agents may be speculative in the sense that they may bid/ask some extremely high/low unit prices to make their bids/asks more likely to be accepted by the auction, and hence, have better chances to earn the payoff associated with the cleared bids. 
Even so, it is not unreasonable to assume that $\mathcal{P}_i$ is bounded. As an MAB game is for decision-making with discrete choices, we also assume that $\mathcal{P}_i$ is countable, which is not restrictive either as prices can at most be of one cent increment. Each discrete unit price in $\mathcal{P}_i$ is considered an `arm' that an agent can choose to bid/ask into an auction. 

\subsection{Rewards}
\label{subsec:rwd_normal}
To facilitate the model setup, we need to introduce the definition of normalized rewards, which take values between 0 and 1. To do so, we first define the benchmark rewards. Let $q_i$ denote the demand/supply of agent $i$, with $q_i < 0 $ for buyers (i.e. $i \in \mathcal{A}_b$), and $q_i > 0$ for sellers (i.e., $i \in \mathcal{A}_s$). For a buyer, we define the lower and upper benchmarks as buying all of $q_i$ at $\overline{P}$ and $P_{FIT}$, respectively. In the opposite, a seller agent has its lower and upper benchmarks defined as selling $q_i$ at $P_{FIT}$ and $\overline{P}$. Using an indicator function, such as $\mathbbm{1}_{\{ i \in \mathcal{A}_b\}}$ to denote that agent $i$ is a buyer if $\mathbbm{1}_{\{ i \in \mathcal{A}_b\}}=1$, and a seller if it is 0, we can write out the uniform mathematical formulas of the lower benchmark reward (denoted by $\underline{\Lambda_i}$) and upper benchmark reward (denoted by $\overline{\Lambda_i}$)  as  
\begin{align}
 & \underline{\Lambda_i} = q_i \cdot [\overline{P} \cdot \mathbbm{1}_{\{ i \in \mathcal{A}_b\}} + P_{FIT} \cdot \mathbbm{1}_{\{ i \in \mathcal{A}_s\}}],\ \mathrm{and} \label{eq:rwd_low} \\[4pt]
& \overline{\Lambda_i} = q_i \cdot [P_{FIT} \cdot \mathbbm{1}_{\{ i \in \mathcal{A}_b\}} + \overline{P} \cdot \mathbbm{1}_{\{ i \in \mathcal{A}_s\}}]. \label{eq:rwd_high}
\end{align}

In an auction setting, the participants send/receive payments based on the cleared results. Specifically, each agent's sent/received payment is calculated according to the agent's cleared price, $p_i^{au}$, and cleared quantity, $q_i^{au}$, as below
\begin{equation}
\label{eq:AuctionPayoff}
\Lambda_i^{au} = p_i^{au}\cdot q_i^{au},\ i\in\mathcal{A}.
\end{equation} 
In \eqref{eq:AuctionPayoff}, the cleared price $p_i^{au}$ is always nonnegative, 
while $q_i^{au}$ can be positive or negative, the same way as how $q_i$ is defined earlier. 
Hence, $\Lambda_i^{au}$ can be positive (for sellers) or negative (for buyers). 
Note that in auctions like the uniform-price double-side auction, there is only one market clearing price. While in some other auctions, such as the maximum volume matching auction to be introduced in Section \ref{subsec:MVM}, it is pay-as-bid for buyers, and receive-as-asked for sellers. Hence, each agent's cleared price $p_i^{au}$ may be different.  

Even in a P2P energy trading market, it is not necessary (or possible) for all agents to buy/sell energy in the P2P market, as some agents' bids/asks may not be (fully) cleared by the market, or there is over supply or over demand in the market. For the uncleared demand, we assume that the corresponding buyers buy from the utility at the rate $\overline{P}$, and for uncleared supply, 
the corresponding sellers sell to the utility at the FIT $P_{FIT}$. We denote the sent/received payment to/from the utility for agent $i$ by $\Lambda_i^{ut}$, and its value is
\begin{equation}
\Lambda_i^{ut} = p_i^{ut}\cdot q_i^{ut},
\end{equation}
where $p_i^{ut} = P_{FIT}$ if $i \in \mathcal{A}_s$ and $p_i^{ut} = \overline{P}$ if $i \in \mathcal{A}_b$, and $q_i^{ut}$ denotes the uncleared quantity.
The total sent/received payment of each agent $\forall i \in \mathcal{A}$ (denoted by $\Lambda_i$) 
is then the sum of $\Lambda_i^{au}$ and $\Lambda_i^{ut}$:
\begin{equation}
\label{eq:rwd_origin}
\Lambda_i = \Lambda_i^{au} + \Lambda_i^{ut}.
\end{equation}

When agent $i$'s auction cleared price $p_i^{au}$ is in $[P_{FIT}, \overline{P}]$, it is easy to see that  $\Lambda_i \in [\underline{\Lambda_i}, \overline{\Lambda_i}]$, and we can define the normalized reward $\pi_i$ as follows: 
\begin{equation}
\label{eq:rwd_normal}
\pi_i = ( \Lambda_i - \underline{\Lambda_i} ) / ( \overline{\Lambda_i} - \underline{\Lambda_i} ), 
\end{equation}
In Eq. (\ref{eq:rwd_normal}), clearly $\pi_i \in [0, 1]$. 
If $p_i^{au} = P_{FIT}$, then a buyer has $\pi_i = 1$  
since $\Lambda_i = \overline{\Lambda_i}$; while a seller has $\pi_i = 0$. If $p_i^{au} = \overline{P}$, then the opposite is true. 
However, as we mentioned earlier, agents may bid/ask prices that are outside the range of $[P_{FIT}, \overline{P}]$. Though it is counter-intuitive, the auction cleared price $p_i^{au}$ could also be outside $[P_{FIT}, \overline{P}]$. In the case that $p_i^{au} < P_{FIT}$, we let $\pi_i = 1 | _{i \in \mathcal{A}_b}$ and  $\pi_i = 0 | _{i \in \mathcal{A}_s}$; for $p_i^{au} > \overline{P}$, we let $\pi_i = 0 | _{i \in \mathcal{A}_b}$ and $\pi_i = 1 | _{i \in \mathcal{A}_s}$. Together with Eq. (\ref{eq:rwd_normal}), we have that 
\begin{equation}
\label{eq:rwd_function}
\pi_i=
\begin{cases}
1 \cdot \mathbbm{1}_{\{ i \in \mathcal{A}_b\}} + 0 \cdot \mathbbm{1}_{\{ i \in \mathcal{A}_s\}}, & \text{for } p_i^{au} < P_{FIT}\\
( \Lambda_i - \underline{\Lambda_i} ) / ( \overline{\Lambda_i} - \underline{\Lambda_i} ), & \text{for }  P_{FIT} \le p_i^{au} \le \overline{P}\\
0 \cdot \mathbbm{1}_{\{ i \in \mathcal{A}_b\}} + 1 \cdot \mathbbm{1}_{\{ i \in \mathcal{A}_s\}}, & \text{for } p_i^{au} > \overline{P}, 
\end{cases}
\end{equation}
where $\underline{\Lambda_i}$, $\overline{\Lambda_i}$, and $\Lambda_i$ are as defined in Eq. (\ref{eq:rwd_low}), (\ref{eq:rwd_high}), and (\ref{eq:rwd_origin}), respectively. 
Converting the actual reward an agent receives/pays in a double-side auction, 
which can be positive or negative, into the normalize reward, which is always in [0, 1] for both buyers and sellers, can greatly simplify both the model setup for the MAB-game and the bidding algorithm design.  

\subsection{Pricing by Bandit Learning}
\label{subsec:pricing_bandit_lrn}
As in Eq. (\ref{eq:rwd_function}), we can see that agent $i$'s reward $\pi_i$ highly depends on its cleared price in an auction, which further depends on the collective actions of the other bidders/sellers. In addition, the same auction is being repeated indefinitely, 
but the set of buyers/sellers may change from rounds to rounds. 
Such a situation is an instance of a dynamic game of incomplete information, 
one of the hardest subject in game theory. 
Incomplete information means that the players do not know the other players' payoff functions or feasible action spaces, nor do they know how many players are in the game. 
Also in a general dynamic game, as opposed to in a so-called repeated game, 
the set of players may change (randomly) over time. 
The standard equilibrium concept for dynamic games of incomplete information is Perfect Bayesian Nash equilibrium (PBNE) \cite{fudenberg1991game, fudenberg1998theory}. 
A PBNE consists of the collection of each player's strategy profile, which
is a function that maps the entire history of the game to
each player's feasible set of actions, under the assumption
that each player updates their beliefs of other players' payoff
functions based on the Bayes' rule. 
Not only the assumptions of PBNE are too strong, as it requires each agent 
to maximize expected payoff (given that other players choosing their corresponding PBNE strategy) over all possible histories of the game, or is such an equilibrium computable, as it would require to find the best strategy profile, which is a function mapping, over the functional space of all possible mappings, leading to an infinite-dimension optimization problem.

To avoid the technical difficulties associated with PBNE, most of the the strong assumptions need to be relaxed. Specifically, we may want to relax the Bayes' updating assumption, and allow consumers to just use a ``good enough" strategy, instead of using the best possible strategy. One way to quantify a ``good enough" strategy is to use the concept of regret, which measures the cumulative differences between what the best action would be in hindsight and what the agent's chosen action was in the past round of the game. 
The various MAB learning algorithms in literature are exactly designed to help an agent choose actions in such a dynamic setting, with provable upper bounds on the cumulative regret. Hence, when the number of rounds of the decision-making goes to infinity, the cumulative regret approaches to 0. However, almost all such algorithms are designed for a single agent, and an underlying assumption for proving bounds on cumulative regret is that the reward distribution associated with each arm (i.e., each feasible action from a finite, discrete space), though unknown to the agent, is stationary. Such an assumption is clearly not true in a multi-agent game, as each arm's reward distribution depends on collectively what the agents do. This lack of stationarity may be the major reason that why there has been very little work on multi-agent MAB games. However, a recent breakthrough on MAB games in \cite{gummadi2016mean} proposes a new concept for such a game with a large number of agents, referred to as a mean-field steady state (MFSS). While it is not a Nash equilibrium in general, 
in the sense that individual agents may have incentives to deviate from their current strategy profile, the keen idea is that when the number of agent is large, the population profile, defined as the histogram of the arm choices of all agents, becomes stabilize as the game continues. Consequently, each arm's reward distribution indeed becomes stationary. Also in \cite{gummadi2016mean}, it is shown that a MFSS exists and is unique under mild conditions. 

The game setting in \cite{gummadi2016mean} is nonetheless very simple. Here we show that the MAB-game framework can indeed be applied to the complex setting of a double-side auction. Specifically, assume that there are $K$ arms in each of agent $i$'s action space $\mathcal{P}_i$. Let $f(k)$, $k=1, \ldots, K$ denote the percentage of the agents that choose arm $k$ in a particular period $h$ of the auction, and we refer 
$\boldsymbol{f} = (f(1), \ldots, f(K))$ as the population profile. 
Then in period $h$, agent $i$'s normalized reward $\pi_i$ (as defined in \eqref{eq:rwd_function}) is a function of $\boldsymbol{f}$ 
and his/her own choice $k$. Agent $i$'s optimal reward in $h$ can be expressed as  
\begin{equation}
\pi_i^*(\boldsymbol{f}) 
= 
\max_{k \in \mathcal{P}_i} \mathbbm{E} [\pi_i( \boldsymbol{f}, k)].
\end{equation} 
Suppose that for the trading-period $h$ across $D$ days, agent $i$ uses a policy $\sigma$, which is an algorithm (aka a mapping) that chooses an arm from $\mathcal{P}_i$ for the next time period based on agent $i$'s own payoff history.\footnote{This is a key difference from PBNE, in which a strategy profile needs to map the entire history of the game, including the past payoffs or actions of \emph{all} the players.} Though the underling optimal reward $\pi_i^*(\boldsymbol{f})$ is unknown to the agent, a policy $\sigma$ can enable the agent to learn about the distributions of rewards for each price arm. Let $\Gamma_{\sigma}(D, k)$ be the number of times price arm $k$ has been chosen by the policy $\sigma$ during all the $D$ rounds. We define agent $i$'s cumulative regret under the policy $\sigma$ for every $D$ rounds as follows:
\begin{equation}
\label{eq:cum_regret}
\Delta_{\sigma} 
= 
\pi_i^*(\boldsymbol{f}) \cdot D 
- 
\sum_{k \in \mathcal{P}_i} \mathbbm{E} [\pi_i(\boldsymbol{f}, k) \cdot \Gamma_{\sigma}(D, k)].
\end{equation}
The regret $\Delta_{\sigma}$ in Eq. (\ref{eq:cum_regret}) is the expected loss due to the fact that the policy does not necessarily always choose the optimal price arm under the population profile $\boldsymbol{f}$, which is unknown to the agent. The policy $\sigma$ is a \textit{no-regret} bandit learning policy if the regret in  Eq. (\ref{eq:cum_regret}) satisfies:
\begin{equation}
\frac{1}{D}\Delta_{\sigma}  < R(D, K),
\end{equation}
for some function $R = o(D)$. For the bandit learning algorithms based on UCB \cite{auer2002finite}, such as UCB1, UCB-tuned and UCB2, it is known that they can yield logarithmic regret bounds; that is, $R(D, K) = \alpha(K) \cdot \frac{1}{D} ln(D)$.

\section{Double Auction Designs}
\label{sec:double_auction}
In this section, we first define the individual monetary utility, corresponding total social welfare, and auctioneer's profit with a P2P energy market auction. Then we discuss three different double-side auction designs that can be applied to clear a P2P market: the uniform-price auction, a variant of Vickrey double-side auction \cite{huang2002design}, and the maximum volume matching auction \cite{niu2013maximizing}. 

\subsection{Social Welfare and Auctioneer's Profit}
As mentioned previously, energy consumers rarely know the precise valuation of the utility associated with their energy production or consumption. To define agents' individual monetary utility, we consider it to be the profit for energy sellers and cost reduction for buyers who participate the P2P market. Since for renewable DER owners, their marginal production costs are zero, the total profit of energy seller $i \in \mathcal{A}_s$ is then 
\begin{equation}
u_i | _{i \in \mathcal{A}_s} =  p_i^{au}\cdot q_i^{au} +  P_{FIT} \cdot q_i^{ut}, 
\end{equation}
which has the same value as $\Lambda_i$ in Eq. (\ref{eq:rwd_origin}). For consumers, they have to pay at $\overline{P}$ without a P2P market, thus we have the cost reduction as
\begin{equation}
u_i | _{i \in \mathcal{A}_b} = (\overline{P} - p_i^{au}) \cdot | q_i^{au}|.
\end{equation}
We refer to the summation of all agents' utility as total social welfare; i.e. $U_{\mathcal{A}} = \sum_{i \in \mathcal{A}} u_i$. 

For the auctioneer (which can be played by the utility or a so-called distribution system operator (DSO)), the total auction trading surplus (denoted by $U_{\mathcal{M}}$) is the sum of bid-ask price difference for each energy unit traded in the auction, which is calculated as below
\begin{equation}
\label{eq:auctioneer_profit}
U_{\mathcal{M}} 
= 
\sum_{i \in \mathcal{A}_b} (p_i^{au}\cdot | q_i^{au} | )
- 
\sum_{i \in \mathcal{A}_s} (p_i^{au}\cdot q_i^{au} ).
\end{equation}

\subsection{Uniform-Price Double-Side Auction}
As plotted in Fig. \ref{fig:uniform_price},  the intersection $( P^*, Q^*)$ of the supply and demand curves determine the market clearing price (and quantity) in a uniform-price double-side auction. All agents pay/receive at the uniform price $P^*$, and a total of $Q^*$ units of energy are traded in the auction. 
It is assumed that the uncleared supply, $\max(Q_s - Q^*, 0)$, is sold to a utility at $P_{FIT}$, in which $Q_s= \sum_{i\in \mathcal{A}_s} q_i$ is the total energy supplied by DERs. Similarly, uncleared demand, if any, is purchased from the utility at $\overline{P}$. Therefore, in Fig. \ref{fig:uniform_price}, the shaded area in light purple represents the total social welfare $U_{\mathcal{A}}$, i.e.
\begin{equation}
U_{\mathcal{A}} = \overline{P} \cdot Q^* + P_{FIT} \cdot \max(Q_s - Q^*, 0).
\end{equation}
Since $p_i^{au} = P^*$ for all agents $i \in \mathcal{A}$, and both $\sum_{i \in \mathcal{A}_b} |q_i^{au}|$ and $\sum_{i \in \mathcal{A}_s} q_i^{au}$ are equal to $Q^*$, by Eq. (\ref{eq:auctioneer_profit}) the auctioneer earns zero profit; i.e. $U_{\mathcal{M}} = 0$.

\begin{figure}[t]
\begin{center}
\subfloat{\includegraphics[width=3.4in]{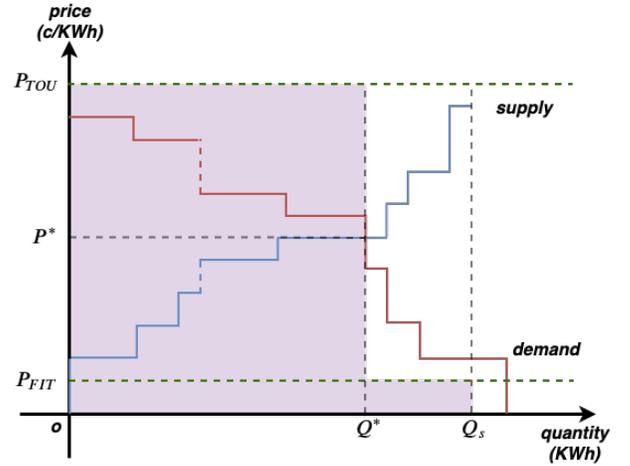}%
	\label{fig:uniform_price}}\\
\end{center}
	\caption{A uniform-price double auction market.}
	\label{fig:uniform_price}
\end{figure}

\begin{figure}[t]
\begin{center}
\subfloat{\includegraphics[width=3.4in]{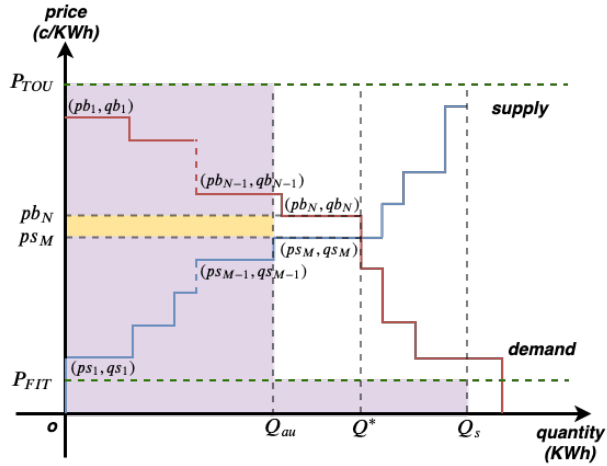}%
	\label{fig:vickrey_like}}\\
\end{center}
	\caption{A Vickrey-like double auction market (Case I).}
	\label{fig:vickrey_like}
\end{figure}

\subsection{Vickrey Variant Double-Side Auction}
Instead of paying/receiving the uniform \textit{equilibrium price}, we consider a Vickrey-like double-side auction as introduced in \cite{huang2002design}. The mechanism works as follows. Similar to the uniform-price auction, all bids/asks are sorted decreasingly/increasingly, and we can have stair-wise demand/supply curves as shown in Fig. \ref{fig:vickrey_like}, in which each stair represents the collective bids/asks at price arm $pb_{n}$/$ps_{m}$. At the critical intersection point $( P^*, Q^*)$ where the aggregate demand and supply meet, there are collective bid $(pb_{N}, qb_{N})$ and ask $(ps_{M}, qs_{M})$.  Then we consider two cases. Case I (as shown in Fig. \ref{fig:vickrey_like}):
\begin{equation}
pb_N \ge ps_M \ge pb_{N+1},
\end{equation}
\begin{equation}
\sum_{m = 1}^{M-1} qs_{m} 
\le
 \sum_{n = 1}^{N} qb_{n}
 \le
 \sum_{m = 1}^{M} qs_{m}
 ,
\end{equation}
and Case II:
\begin{equation}
ps_{M+1} \ge pb_N \ge ps_{M},
\end{equation}
\begin{equation}
\sum_{n = 1}^{N-1} qb_{n} 
\le
 \sum_{m = 1}^{M} qs_{n}
 \le
 \sum_{n = 1}^{N} qb_{m}
 .
\end{equation}

Herein, we describe the clear mechanism for Case I; Case II is similar.
\subsubsection*{\textbf{Rule 1}}
If $\sum_{n=1}^{N-1} qb_n \ge \sum_{m=1}^{M-1} qs_m$, there is overdemand. All the asks with $m < M$ sell all their supply $qs_m$ at price $ps_M$; all the asks with $m \ge M$ sell their supply at $P_{FIT}$ to the utility. All the bids with $n < N$ buy at $pb_N$ and each of them buys a volume equal to $qb_n - (\sum_{n=1}^{N-1} qb_n - \sum_{m=1}^{M-1} qs_m) / (N - 1)$; all the uncleared bids buy at $\overline{P}$ from the utility.

\subsubsection*{\textbf{Rule 2}}
If $\sum_{n=1}^{N-1} qb_n \le \sum_{m=1}^{M-1} qs_m$, there is oversupply. All the bids with $n < N$ buy all their demand $qb_n$ at price $pb_N$; all the bids with $n \ge N$ buy from a utility at $\overline{P}$. All the asks with $m < M$ sell at $ps_M$ and each of them sells a volume equal to $qs_m - (\sum_{m=1}^{M-1} qs_m - \sum_{n=1}^{N-1} qb_n)/ (M - 1)$; all the unsuccessful asks sell at $P_{FIT}$ to the utility.

According to the clear rules, the total trade volume in the auction is 
\begin{equation}
Q_{au} = min(\sum_{n=1}^{N-1} qb_n , \sum_{m=1}^{M-1} qs_m)
.
\end{equation}
Then the total social welfare for all agents can be calculated as below (which is represented by the light purple area in Fig. \ref{fig:vickrey_like})
\begin{equation}
U_{\mathcal{A}}
=
[(\overline{P} - pb_N) + ps_M]\cdot Q_{au}
+
P_{FIT} \cdot (Q_s - Q_{au}) 
.
\end{equation}

The auctioneer's profit represented by the yellow shaded area in Fig. \ref{fig:vickrey_like} is as below
\begin{equation}
U_{\mathcal{M}}
=
(pb_N - ps_M) \cdot Q_{au} 
.
\end{equation}

\subsection{Maximum Volume Matching Double-Side Auction}
\label{subsec:MVM}
Other than aiming to maximize social welfare for agents or profit for the acutionner, the auction design proposed in \cite{niu2013maximizing} is to  maximize the traded volume with given a set of bids and asks. The idea of market clear can be intuitively illustrated in Fig. \ref{fig:max_vol}. Suppose that the demand/supply curves are based on the bids/asks shown in Fig \ref{fig:uniform_price}. The supply curve is flipped horizontally and then shifted right towards the demand curve until the two curves touch. The distance  (denoted by $Q_{au}$) the flipped supply curve can move is the minimal horizontal distance between the flipped supply curve and the demand curve, which is exactly the maximum trading volume achievable by the auction. For energy quantity from 0  to $Q_{au}$, the corresponding bids $(pb_n, qb_n)$ on the demand curve and asks $(ps_m, qs_m)$ on the shifted supply curve are matched, and then successfully matched buyers/sellers pay/receive at their bid/ask price, respectively. Let $\mathcal{S}_b$ and $\mathcal{S}_a$  denote the set of successful bids and asks, respectively. As before, uncleared supply is assumed to be sold to a utility at $P_{FIT}$, and uncleared demand is bought at from the utility at $\overline{P}$.

According to the clearing mechanism, the total social welfare of all agents is as below (represented by the light purple area in Fig. \ref{fig:max_vol}): 
\begin{equation}
U_{\mathcal{A}} 
= 
\sum_{n \in \mathcal{S}_b} (\overline{P}-pb_n) qb_n 
+
\sum_{m \in \mathcal{S}_a} ps_m  qs_m
+
P_{FIT}  (Q_s - Q_{au}) 
.
\end{equation}

The auctioneer's profit is still the auction trading surplus (represented by the yellow shadow area in Fig. \ref{fig:max_vol}) as below:
\begin{equation}
U_{\mathcal{M}} 
= 
\sum_{n \in \mathcal{S}_b} (pb_n \cdot qb_n)
-
\sum_{m \in \mathcal{S}_a} (ps_m \cdot qs_m)
.
\end{equation}


\begin{figure}[t]
\begin{center}
\subfloat{\includegraphics[width=3.6in]{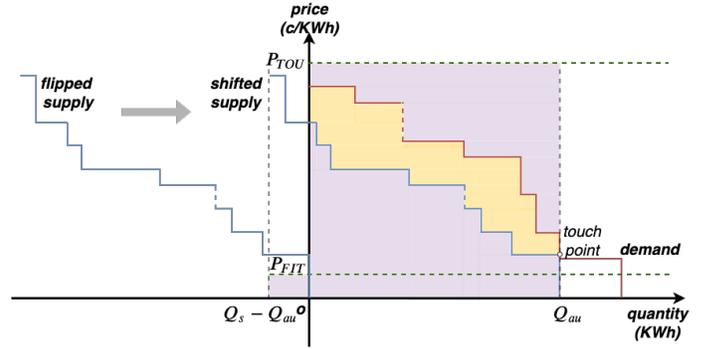}%
	\label{fig:max_vol}}\\
\end{center}
	\caption{A maximum volume matching auction market.}
	\label{fig:max_vol}
\end{figure}

\section{Numerical Simulations}
\label{sec:num_res}
In this section, we apply the MAB-game framework to study the market outcomes of three different auction designs to realize a P2P energy trading market.

\subsection{Input Data}
\label{sec:sim_data}

\subsubsection{Decision epochs and temporal resolution}
As a starting point, we do not consider time-linking constraints in our models, and each trading window is independent of others in a day. The simulations presented herein concern a single one-hour trading period for the peak hour 17:00 - 18:00 across 300 days, i.e. $D=300$. 

\subsubsection{$\overline{P}$, FIT and decision space}
We  choose $\overline{P} = 11$  $\cent$/KWh and $P_{FIT} = 5$ $\cent$/KWh. All agents have the same decision space $\mathcal{P}_i$ that contains all the discretized price arms from $0$ $\cent$/KWh to $14 $ $\cent$/KWh of one cent increment. Hence, each agent has 15 arms (including 0 $\cent$/KWh) to choose from in each round of the auction.  

\subsubsection{Bandit learning algorithms for pricing}
For choosing price arms to bid/ask in the auctions, each agent $i \in \mathcal{A}$ can choose one of the four learning algorithms: UCB1, UCB-tuned, UCB2, and $\epsilon - greedy$, with equal probabilities. Details of such algorithms (for a single agent) can be found in \cite{auer2002finite}.

\subsubsection{Consumers}
In the numerical test cases, we simulate 2000 distributed residential household consumers that participate in the auctions, i.e. $| \mathcal{A}_b | = 2000$. According to the Residential Residential Energy Consumption Survey (RECS) by U.S. Energy Information Administration (EIA) \cite{usres}, a residential customer consumes about 30 KWh per day on average. Considering that it is a peak hour, we let consumers repeatedly sample their energy demand quantities from a \textit{Uniform} distribution $U(1.5,2)$ in KWh, independently, for the same hour across days, which is slightly higher than the average consumption level per hour.

\subsubsection{Prosumers}
On the sell-side, we also consider 2000 prosumers with DERs, i.e. $| \mathcal{A}_s | = 2000$. For the DERs, we only consider two renewable resources, solar and wind, in this work. Due to the popularity of distributed residential solar panels, we assume that $4/5$ of the prosumers have solar-based distrbuted generation; while the other $1/5$ have wind-based. In the simulations, we use the System Advisor Model (SAM) \cite{nrelsam} developed by National Renewable Energy Laboratory (NREL) to model residential generation output by solar and wind. The weather resource data for the state of Arizona from NREL is used as inputs for the SAM model. 

For solar PV generation, we assume that all panels have nameplate capacity of 2 KWdc, with DC to AC ratio of 1.2 and inverter efficiency of 96\%. For each solar PV owner, the module type and array type have equal chance to be one of \{\textit{Standard, Premium, Thin Film}\} and \{\textit{Fixed Open Rack, Fixed Root Mount, 1 Axis Tracking, 1 Axis Backtracking, 2 Axis Tracking}\}, respectively. All other inputs are set as default in the \textit{Photovoltaic PVWatts} simulations for distributed residential in SAM. More details about photovoltaic simulations can be found in \cite{nrelsam,blair2014system}.

For the simulations of distributed residential wind generation, each wind-based prosumer samples its turbine model uniformly from the 8 wind turbine models listed in Table \ref{tb:wind_turbine}, and the number of turbines owned by the prosumer is uniformly sampled among 1 through 4. All other inputs are set as default in the \textit{Wind Residential} simulations in SAM. The turbines' specifications, such as wind power curves and turbine layout, can be found in \cite{nrelsam,blair2014system}.

\begin{table}[t]
\caption{Wind turbine models}
\centering
\begin{tabular}{|c|c|}
\hline
Model                                    & KW Rating             \\ \hline
Energy Ball HEA V100 1.1m 0.6KW & 0.5 \\
Bergey BWC XL.1                          & 1                     \\
True North Power Arrow 2m 1KW            & 1.23                  \\
Future Energy FE1048U 1.8m 1KW           & 1.5                   \\
Hummer 3.1m 1KW                          & 2                     \\
Energy Ball HEA V200 1.98m 2.5KW         & 2.23                  \\
Southwest Windpower Skystream 3.7m 1.9KW & 2.63                  \\ 
Westwind 3.7m 3KW & 3.1\\\hline
\end{tabular}\vspace*{-10pt}
\label{tb:wind_turbine}
\end{table}

\subsection{Numerical Results}
In the following, we use UP, VV, and MV to denote uniform price auction, Vickrey variant auction, and maximum volume matching auction, respectively. 
In Fig. \ref{fig:clear_quant}, the cleared quantities of the auctions are presented, and we can see the trend of convergence to a steady state. A  counter-intuitive observation is that in the later phase, UP is more likely to have a higher level of traded volume than MV, even though MV is designed to maximize traded volume. The reason is that with bandit learning, agents are updating their bids/asks dynamically, and thus the collective bids/asks schedules are not necessarily the same across different auctions. (With the same set of bids/asks, MV will yield the most cleared quantity for sure.) Besides the volume, we can see after a while of learning, UP's total cleared quantity has smaller volatility than the other two auction designs. Therefore, in terms of auction cleared quantity, UP outperforms VV and MV, and thus the auction design can let more renewable DERs be utilized in a P2P market.

Similar to the cleared quantity, agents' total social welfare also shows the trend of convergence, as seen in Fig. \ref{fig:social_welfare}. Associated with more cleared quantity, buyers and sellers in UP have higher social welfare than in the other two auctions in the later auctions. The performance of VV and MV are close to each other. Accordingly, for the total normalized reward, the results display very similar patterns as shown in Fig. \ref{fig:reward_total}.

Though UP outperforms the other two auctions for benefiting market participants and incentivizing DERs, it may not be preferred by the auctioneer. As discussed in Section \ref{sec:double_auction}, the auctioneer earns no profit in UP due to the zero trading surplus. According to simulation results, the auctioneer can achieve the most profit in MV, though the profit fluctuations of MV are much higher than VV's.

\begin{figure}[t]
\begin{center}
\subfloat{\includegraphics[width=3.5in]{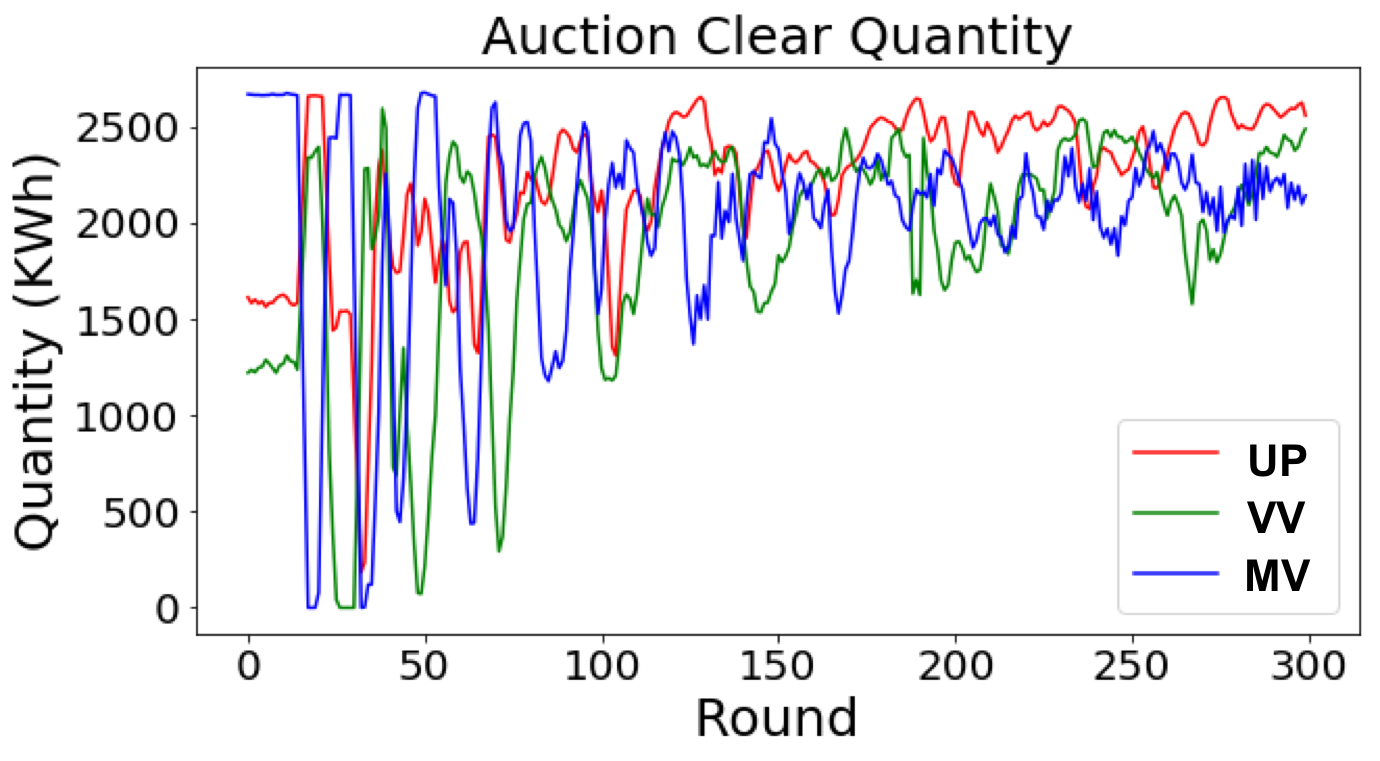}%
	\label{fig:clear_quant}}\\
\end{center}
	\caption{Total clear energy quantities (KWh) in the auctions.}
	\label{fig:clear_quant}
\end{figure}

\begin{figure}[t]
\begin{center}
\subfloat{\includegraphics[width=3.5in]{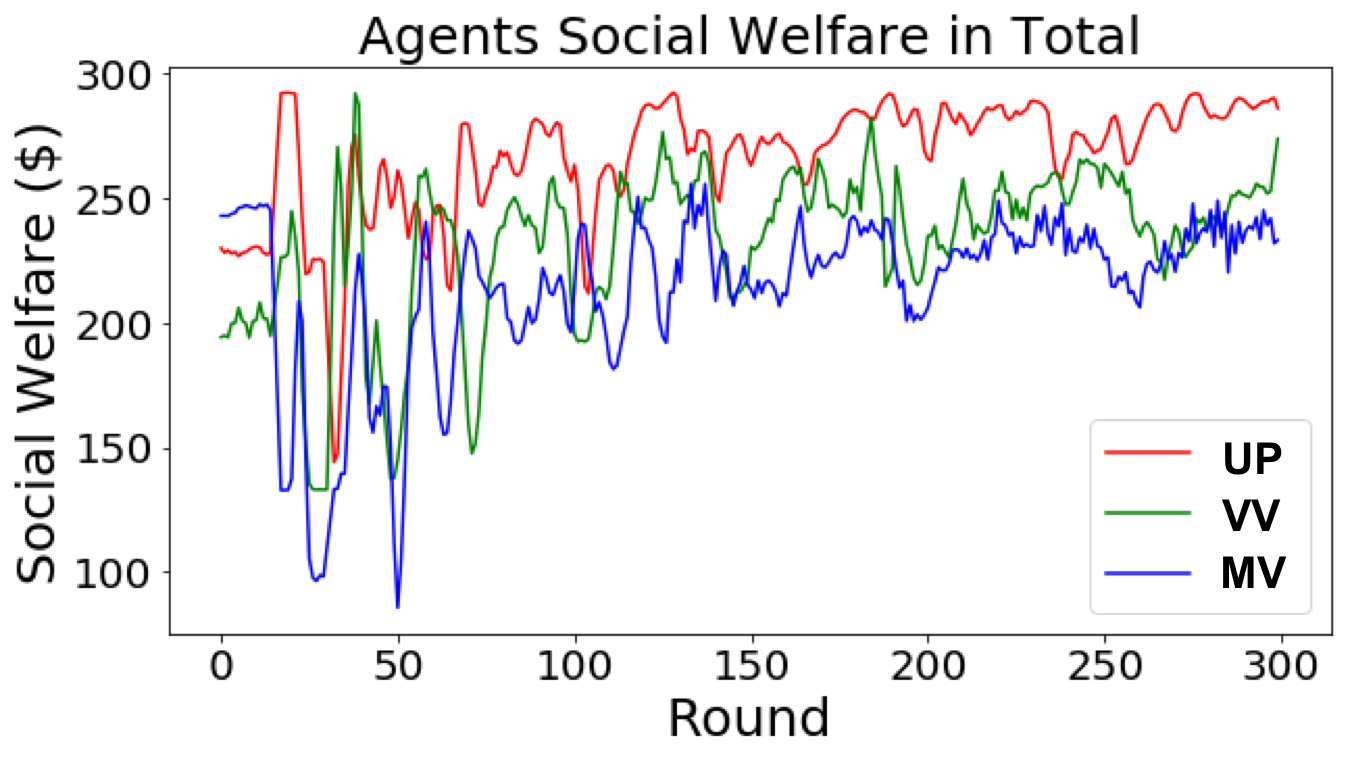}%
	\label{fig:social_welfare}}\\
\end{center}
	\caption{Total social welfare (\$) of all buyers and sellers in the auctions.}
	\label{fig:social_welfare}
\end{figure}

\begin{figure}[t]
\begin{center}
\subfloat{\includegraphics[width=3.5in]{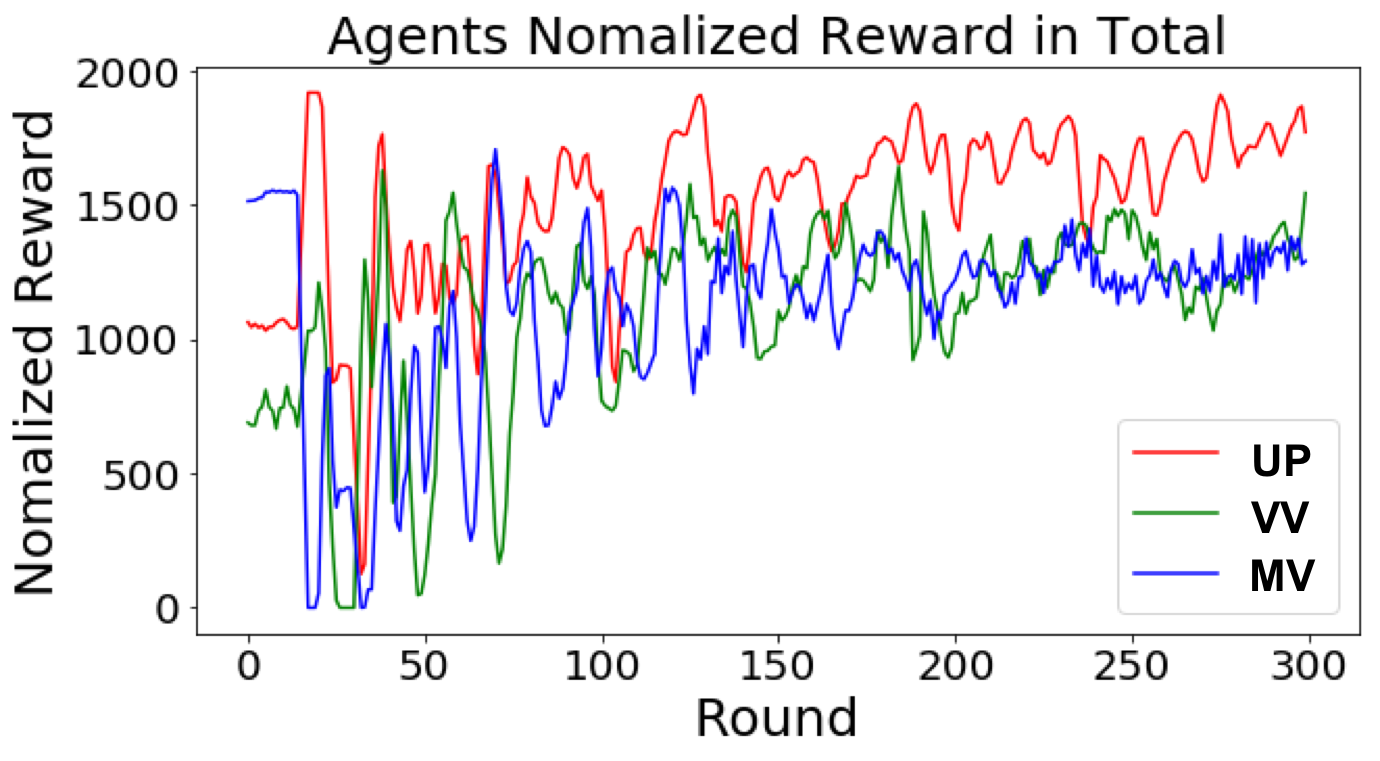}%
	\label{fig:reward_total}}\\
\end{center}
\caption{Total normalized reward of all buyers and sellers in the auctions.}
	\label{fig:reward_total}
\end{figure}

\begin{figure}[t]
\begin{center}
\subfloat{\includegraphics[width=3.5in]{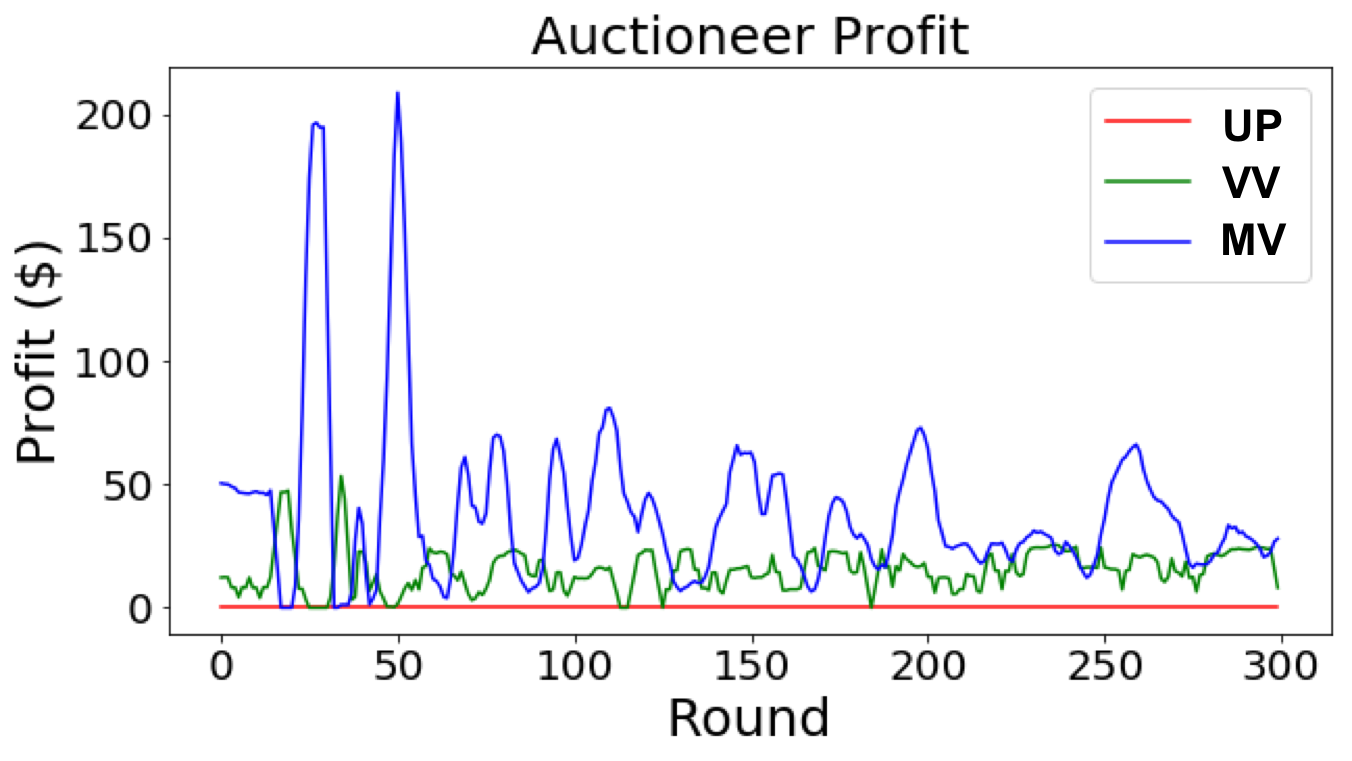}%
	\label{fig:auctioneer_profit}}\\
\end{center}
	\caption{Auctioneer's profit (\$) in the auctions.}
	\label{fig:auctioneer_profit}
\end{figure}

\section{Conclusion}
\label{sec:conclusion}
In this work, we propose a MAB-game approach to help market participants choose prices for their energy bid/ask in a P2P double-side auction. The bandit learning approach allows each agent to make a decision based only on his/her own history, 
which makes this approach implementable in real-world settings. We conduct simulations based on the MAB-game framework under three different auction designs, and the results indicate the convergence of cleared quantities, total social welfare and total normalized reward for agents. Moreover, based on simulation results, the uniform-price double auction appear to outperform the other two in terms of market participants' benefits. For the auctioneer, the maximum volume matching offers the highest profit. For future research directions, a pressing need is to investigate how to incorporate time-linking constraints in the MAB-framework so that energy storage can be explicitly considered. Another need is to verify if the technical conditions in \cite{gummadi2016mean} for the existence and uniqueness of a MFSS are satisfied in the specific game setting here. 
\bibliographystyle{IEEEtran}
\bibliography{paper_ref.bib}

\end{document}